\documentstyle[12pt,epsf]{article}
\catcode`\@=11
parenthesis
ref
\def\beq{\begin{equation}}
\def\eeq{\end{equation}}
\def\beqa{\begin{eqnarray}}
\def\eeqa{\end{eqnarray}}
\def\rd{{\mathrm d}}

\def\intd4x{\int{\rd}^4x}

\def\m32{{m_{3/2}}}

\begin{document}

\begin{titlepage}
\begin{flushright}
FERMILAB-Pub-94/324-A
\end{flushright}

\begin {center}
{\Huge Light mesons production at the Tevatron to next-to-leading  
order}

\end{center}
\vspace{1.5cm}
\begin{center}

{\bf Mario Greco$^{a}$ and Simona Rolli$^{b}$ \footnote{e-mail 

address:greco@lnf.infn.it, rolli@fnalv.fnal.gov}}\\
\vspace{1cm}
{$^a$\it Dipartimento di Fisica, Universit\`a dell'Aquila, Italy}\\
{\it INFN, Laboratori Nazionali di Frascati, Frascati, Italy}\\
\vspace{0.5cm}
{$^b$\it Dipartimento di Fisica Nucleare e Teorica,,Universit\`a di  
Pavia,
Italy}\\
{\it INFN, Sezione di Pavia, Pavia, Italy and}\\
{\it NASA/Fermilab Astrophysics Center, FERMILAB, Batavia, Il  
60510}\\

\end{center}

\vspace{2cm}
\begin{center}
{\bf Abstract}
\end{center}
Inclusive production of light mesons ($\pi^0$, $\eta$, $\pi^{\pm}$,  
$K^{\pm}$ )
at the Tevatron is considered in QCD to next-to-leading order 

in the formalism of fragmentation functions.
We present various distributions of phenomenological interest, along  
with  

a new set of $K$ mesons fragmentation functions.

\vskip1truecm

\vfill
\end{titlepage}

The study of inclusive particle production in hadron-hadron  
collisions is 

interesting for giving information on the fragmentation properties of  
partons 

in perturbative QCD and the hadronization mechanisms.
From the theoretical point of view this means that one has to know  
separately 

parton distribution functions, partonic cross sections and  
fragmentation 

functions: the factorization theorem in fact insures that the  
inclusive cross 

section can be written as a convolution of these different terms. 

While different sets of parton distribution functions do exist at 

next-to-leading order(NLO) \cite{pdf},  and also partonic cross  
sections 

have been 

evaluated to one-loop accuracy \cite{aversa}, for the fragmentation  
functions
( FF) we are only at the beginning, only a few sets existing at NLO  
accuracy
\cite{noi,noieta,fermi,copia,lep}.
%[REFERENZE 3,4,5,6,7]. 

Indeed the study of calorimetric jet cross sections
has been of primary interest during last few years, and only now
experiments are providing us information on inclusive production of  
single 

hadrons.

In previous analyses \cite{noi,noieta,fermi} FF have been evaluated 

at next-to-leading order for neutral and charged pions, and for the  
eta meson, 

using two different methods.
In the first approach, one starts with a very simple parametrization  
of the
parton FF at $Q_0^2 \sim$ few GeV$^2$, and fits $e+e-$ data after NLO  
evolution.
Alternatively, a Montecarlo simulator is used to parametrize the  
parton FF
at $Q_0^2 \sim$ 30 GeV$^2$ and, again after NLO evolution,
$e^+e^-$ data are well reproduced.
Both methods are in good agreament, and indeed with these sets we  
were able 

to reproduce data obtained in hadron collisions in 

fixed target experiments and at the Cern $Sp\bar pS$ collider.

In this Letter we present a new set of FF for charged and neutral  
kaons, 

along with
various distributions of phenomenological interest regarding  the  
production 

of light mesons  at high $p_t$, for
$\sqrt{s}= 1800$ GeV, due to the lack of NLO predictions at these  
energies.
Indeed  in a previous work Borzumati et al. \cite{franc} have given  
some 

predictions on inclusive hadron production at Tevatron, but in a low  
$p_t$ 

range and using obsolete LO sets of fragmentation functions  
\cite{dg}.

We would like to remind that the study of neutral clusters production  
could 

be useful in the evaluation of background to a possible Higgs signal.
On the other hand inclusive 

particle production will help in understanding the jet 

fragmentation properties in QCD

In the following we will first briefly discuss the theoretical  
framework 

for calculating the single 

inclusive particle cross section in hadron-hadron collisions, and 

then we will present and discuss various phenomenological results.

We report, for reader's convenience, the main formulae for one hadron  
inclusive 

production at
next-to-leading-order, via the  reaction $ A+B\to H
+ X$, 

where A
and B are incoming hadrons. The cross section is given by
the convolution of the partonic cross section and the fragmentation
functions $D_l^H(z,M_f^2)$: 

\begin{equation}
E_{H} \; \frac{d\sigma_{A+B \rightarrow H}}{d^3 \vec{P}_{H}}
 =  \sum_{l} \; \int_{z_H}^{1} \frac{dz}{z^2} \; D^{H}_{l}(z,
M^2_f) \; E_l \: \frac{d\sigma_{A+B \rightarrow l}}{d^3 \vec{P}_{l}}
(\frac{z_H}{z}, \theta, \alpha_s(\mu^2), M^2_f,\cdots) ,
\label{meq}
\end{equation}
where $z_H$ is the reduced energy of the hadron $H$,
$z_H=2 E_H / \sqrt S$, 

$\theta$ is the scattered angle of the parton l, and
the inclusive production of the parton l in the reaction
$A+B \rightarrow l$
has the following perturbative development:
\begin{equation}
E_l \; \frac{d\sigma_{A+B \rightarrow l}}{d^3 \vec{P}_{l}}
(\frac{z_H}{z}, \theta, \alpha_s(\mu^2), M^2_f,\cdots ) =
\sigma^0_{A+B \rightarrow l}(\frac{z_H}{z},\theta) + \frac{\alpha_s
(\mu^2)}
{2 \pi} \; \sigma^1_{A+B \rightarrow l}(\frac{z_H}{z},\theta,M^2_f) +
\cdots .
\end{equation}
$D^H_l(z,M^2_f)$ represents the number of hadrons H
inside the parton l carrying the fraction of momentum $z$ from H,
evolved at the scale $M^2_f$.
These fragmentation functions satisfy the usual Altarelli-Parisi type
evolution equations .

The partonic cross-sections are \cite{aversa,noi}:
\begin{eqnarray}
E_l \frac{d\sigma_{p+p \rightarrow l}}{d^3 \vec{P}_{l}}
(y,\theta,\alpha_s(\mu^2),M_f^2) & = &
\frac{1}{\pi S} \sum_{i,j} \int_{V W}^{V} \frac{dv}{1-v}
\int_{V W/v}^{1} \frac{dw}{w} \nonumber \\
& & \mbox{} \times \left[ F^p_i(x_1,M^2)
F^p_j(x_2,M^2) \left( \frac{1}{v} \left( \frac{d \sigma^0}{dv}
\right)_{i j \rightarrow l} (s,v) \delta(1-w) \right. \right.  
\nonumber
\\
& & \left. \left. + \frac{\alpha_s(\mu^2)}{2 \pi} K_{i j \rightarrow  
l}
(s,v,w;\mu^2;M^2,M_f^2) \right) + (x_1 \leftrightarrow x_2) \right] ,
\end{eqnarray}
where the partonic variables are $s=x_1x_2S$ and 

\[ x_1= \frac{VW}{vw} , \; \; x_2  = \frac{1-V}{1-v}, \]  

and the hadronic ones are defined by:
\[ V = 1- \frac{y}{2} (1-\cos \theta), \; \; W= \frac{y (1+\cos  
\theta)}
{2 - y (1-\cos \theta)} . \]

In the partonic cross sections to one loop \cite{aversa},
calculated from the squared matrix elements
$O(\alpha_S^3)$ of Ellis et Sexton \cite{ellis}, the initial 

state collinear divergences
have been factorised and absorbed into the dressed structure  
functions in
the $\overline{MS}$ scheme. 

Coherently with this choice, we have used for the proton structure  
functions
 set B1 of Morfin \& Tung, 

\cite{mt} (set A), set  MRS D0 of Martins Roberts \& Stirling  
\cite{mrs}
(set B),      

and set GRV HO of Gl\"uck, Reya \& Vogt \cite{grvpro} (set C).

We have used $\alpha_s$ calculated at 2-loop, with 5 flavours and  
with 

$\Lambda_{QCD} =~200~MeV$. Set A of nucleon structure functions has  
been 

indeed evolved with $\Lambda_{QCD} = ~194~MeV$.

As already stated above, FF will be considerd to NLO accuracy. For  
the 

$\pi^0$
case, various consistent parametrizations have been discussed in 

ref.~\cite{noi},
using different methods and initial conditions. All of them have been
successfully compared with the current experimental data in $e^+ e^-$  
and 

$p\bar p$
collisions at various energies. In the following we will use only one  
set 

of them \cite{fermi},
% [REFERENZA ULTIMA PARAMETRIZZAZIONE Z.PHYS.C]
based on the latter version of the MonteCarlo simulator HERWIG 

\cite{herwig,herwig2}, which is used to fix the 

initial conditions at the fragmentation scale $M_0$ = 30 GeV. The  
same method 

has also been applied in ref.\cite{noieta} to inclusive $\eta$  
production, and
indeed the predicted
$\eta/\pi^0$ ratio has been found to agree with the present  
experimental
information at ISR \cite{isr} and from $e^+ e^-$ and $p\bar p$  
colliders. 

Recent 

fixed target experiments also agree \cite{e706} with the predictions  
of ref.
 \cite{noieta}, and further constrain the gluon FF for $\eta$.
In reference \cite{fermi} we have adopted the same technique to  
extract the
charged pion FF, in good agreement with the neutral pion set assuming
$SU(2)$ symmetry.

We are therefore quite confident of the reliability of the method and 

consequently we use the same tecnique to obtain a new set of FF
for charged kaons  and $K^0_s$: they are very similar in 

slope (with the appropriate interchanges between  $u$, $\bar u$ and  
$d$, $\bar d$), 

%($u, \bar u$ and $d, \bar d$ COSA VUOI DIRE?), 

differing only by a
normalization factor, as we expect. So we will present final results  
for
for the production cross sections for charged kaons only.

As usual the functions are parametrized as:
\beq
D^{h}_i(z, M_0^2)=N_iz^{\alpha_i}(1-z)^{\beta_i}
\eeq
where $i$ runs over $(u,~d,~s,~c,~b,~g)$, and $M_0$ = 30 GeV.
The coefficients are given
in Table I and II (we indicate the average multiplicity of produced  
hadrons with
the symbol $<n_i>$).

We now present various numerical results, starting, in Fig.1, with  
the $p_t$ 

distribution 

for inclusive single particle production for 

$\pi^0$, $\eta$, $\pi^{\pm}$, $K^{\pm}$, integrated in the region of 

pseudorapidity $\eta=-\ln(\tan({\theta\over 2}))$ between -0.7 and  
0.7 and using 

the Set B-1 of Morfin and Tung \cite{mt}. We set all the scales equal  
to 

the $p_T$ of the produced hadron.

In Figs. 2 we study the dependence on factorization, fragmentation
and normalization scales. We vary the scales simultaneusly for  

different values of $p_t$ and obtain that the cross sections change  
of
about $20\%$ for $0.5 < \xi < 2$, the variation being reduced at 

high $p_T$. This is explicitly shown in Figs. 3, where we plot the
$p_t$ distributions  for different choices of the scale  
($\xi=M/p_t$=0.5,
1, 2).

In Figs. 4 we present the $\eta$ distributions for $\pi^0$ and  
$\pi^{\pm}$ 

production for different values of $p_t$ in the region $|\eta| \le  
2$.

In Figs. 5 we show the contributions from different subprocesses, 

i.e. $qq$, $qg$ and $gg$, where $q$ refers to the sum of $q$ and  
$\bar q$ in the initial state,
 for 

$\eta=0$ and using  Set B1 of Morfin and Tung. 

In order to study the sensitivity to different parton distributions  
we plot
in Figs. 6 the ratio of the cross sections for
the subprocess $gg \to h + X$ for two other sets of proton structure  
functions,
 i.e. MRS-D0,  GRV-HO, with respect to the set MT-B1. The variations  
are 

confined within about 10\%. The sum over all parton subprocesses  
generally
reduces the effect, as shown 

in Figs. 7, with the exception of the GRV-HO set for kaons  
production.

In order to disentangle the fragmentation properties and the  
hadronization 

mechanism of high $p_t$ jets, we consider next the ratio between 

the single hadron and jet cross sections, for fixed values of the  
variable
$z=E_{hadr}/E_{jet}$. Then, using the jet algorithm of  
ref.\cite{jets,huth}
and the NLO evaluation of the jet cross sections of
ref. \cite{aversa},  we present   

in Fig.8 a preliminary result\cite{work} on jet fragmentation in 

charged and neutral pions, with the energy of the jet varying between  
40 and 

70 GeV, with a jet cone radius R=0.7 centered around the $\eta=0$  
direction.
The overall theoretical uncertainty -which is not reported in figure- 

can be estimated to be of order 50\%. 

%[ERRORI TH?] 

We also show the analogous
experimental result on jet fragmentation in charged  
hadrons\cite{hubbard}, in reasonable
agreement with the theoretical prediction.

As a final result we give a mean value for the  ratio R=$\eta/\pi^0=  
1.15\pm .30 $ , in the 

range 20 Gev  $< p_t <$  200 GeV  which agrees with the  recent 

experimental value of R=$1.02\pm 0.15 \pm 0.23$ \cite{diph}. The  
theoretical uncertainty is related to the variation of the scales in  
parton distributions and fragmentation functions.

\vspace{1cm}

To conclude, we have presented various theoretical results
 of phenomenological interest at the Tevatron, concerning the 

inclusive  production of light mesons in QCD to NLO.
 We have studied the uncertainty with respect to the choice of the  
scales and 

different sets of proton structure functions. 

We have also presented new sets of fragmentation functions for
neutral and charged kaons.

One of us (S.R.) would like to aknowledge Joey Huston and Rob Blair  
of CDF Collaboration for useful discussions, and Andrea Parri of CDF  
and KLOE collaborations for precious suggestions about graphics.

This work was supported in part by the DOE and by the NASA  
(NAGW-2831) at
Fermilab.
\vfill\eject
{\bf Table Captions}

\begin{itemize}
\item{Table I:} parameters of the $K^{\pm}$ fragmentation functions 

at $M_0=30$ GeV (see eq.4).
\item{Table II:} parameters of the $K^0_s$ fragmentation functions 

 at $M_0=30$ GeV (see eq.4).
\end{itemize}
\vspace{0.5cm}

{\bf Figure Captions}
\begin{itemize}
\item{Fig. 1:} $P_t$ distribution for $\pi^0$, $\eta$, $\pi^{\pm}$, 

$K^{\pm}$ production, at 

$\sqrt{s}=1800$ GeV and $|\eta|\le 0.7$.

\item{Figs. 2:} $\pi^0$, $\eta$, $\pi^{\pm}$, $K^{\pm}$ production, 

$\sqrt{s}=1800$ GeV. 

Dependence of ${{d\sigma}\over{ dp_t}}$ on
the renormalization, factorization and fragmentation mass scales:  

$\mu=M_p=M_f=\xi p_t$ 

for $p_t=20,~80,~160,~200$ GeV and $|\eta|\le 0.7$.

\item{Figs. 3:} $P_t$ distribution for $\pi^0$, $\eta$, $\pi^{\pm}$,  
$K^{\pm}$ 

production, 

$\sqrt{s}=1800$ GeV . 

for different values of $\xi=M/p_t$, and using Set B1 of Morfin \&  
Tung.

\item{Figs. 4:} $\pi^0$, $\pi^{\pm}$ production.
$\eta$ distributions of 

${{d\sigma}\over{d\eta dp_t}}$ at fixed $p_t$.

\item{Figs. 5:} $\pi^0$, $\eta$, $\pi^{\pm}$, $K^{\pm}$ production, 

$\sqrt{s}=1800$ GeV.
%$p_t$ distributions of 

${{d\sigma}\over{d\eta dp_t}}$, for various partonic
subprocesses ( $q$ refers to the sum of $q$ and $\bar q$ in the  
initial state).

\item{Figs. 6:} $\pi^0$, $\eta$, $\pi^{\pm}$, $K^{\pm}$ production, 

$\sqrt{s}=1800$ GeV.
$p_t$ distribution  

at $\eta=0$ , for the subprocess
$g g\to \pi^0 + X$ for different sets of proton structure
functions. Curves are normalized to the MT-B1 set of proton structure  
functions.

\item{Figs. 7:}$\pi^0$, $\eta$, $\pi^{\pm}$, $K^{\pm}$ production, 

$\sqrt{s}=1800$ GeV. Dependence of
${{d\sigma}\over{d\eta dp_t}}$ on different sets of parton  
distributions. 

Curves are normalized to the set MT-B1 of proton structure functions.

\item{Fig.8:} jet fragmentation function, into $\pi^0$ and $(\pi^+ +  
\pi^-)$.
Ratio of the inclusive hadron cross section to the inclusive jet  
cross section
for $h=\pi^0$ and $(\pi^+ + \pi^-$), as function of $z =  
E_h/E_{jet}$. The experimental 

points refer to charged hadrons and are from reference  
\cite{hubbard}.               

\end{itemize}

%%%%%%%%%%%%%%%%%%%%%%
\vfill\eject
$$
\begin{tabular}{|c|c|c|c|c|}
\hline
%\multicolumn{4}{|c||}{$\delta = 0.35$}
%&\multicolumn{4}{c|}{$\delta = 0.40$}\\
\hline
\multicolumn{1}{|c|}{$Parton$} &
\multicolumn{1}{c|}{$\alpha$} & \multicolumn{1}{c|}{$\beta$}
&\multicolumn{1}{c|}{$N_i$}
&\multicolumn{1}{c|}{$<n_i>$}\\ 

\hline
u & $-1.42 \pm 0.03$ & $1.48 \pm 0.13$ & 0.1 & 0.59  \\ \hline
d & $-1.10 \pm 0.03$ & $4.32 \pm 0.05$ & 0.34 & 0.55  \\ \hline
s & $-0.83 \pm 0.03$ & $1.13 \pm 0.03$ & 0.57 & 0.95  \\ \hline
c & $-0.70 \pm 0.03$ & $3.78 \pm 0.08$ & 1.41  & 1.01  \\ \hline
b & $-0.77 \pm 0.03$ & $7.7 \pm 0.18$ & 2.82 & 1.24  \\ \hline
g & $-0.39 \pm 0.03$ & $4.74 \pm 0.08$ & 1.97 & 0.62  \\ \hline
\end{tabular}
$$
\centerline{ Table I}
\vspace{0.2cm}
%%%%%%%%%%%%%%%%%%%%%%
$$
\begin{tabular}{|c|c|c|c|c|}
\hline
%\multicolumn{4}{|c||}{$\delta = 0.35$}
%&\multicolumn{4}{c|}{$\delta = 0.40$}\\
\hline
\multicolumn{1}{|c|}{$Parton$} &
\multicolumn{1}{c|}{$\alpha$} & \multicolumn{1}{c|}{$\beta$}
&\multicolumn{1}{c|}{$N_i$}
&\multicolumn{1}{c|}{$<n_i>$}\\ 

\hline
u & $-1.06 \pm 0.03$ & $4.37 \pm 0.13$ & 0.19 & 0.27  \\ \hline
d & $-1.39 \pm 0.03$ & $1.46 \pm 0.05$ & 0.05 & 0.28  \\ \hline
s & $-0.84 \pm 0.03$ & $1.01 \pm 0.03$ & 0.26 & 0.45  \\ \hline
c & $-0.80 \pm 0.03$ & $3.31 \pm 0.08$ & 0.50 & 0.49  \\ \hline
b & $-0.63 \pm 0.03$ & $8.15 \pm 0.18$ & 2.02 & 0.62  \\ \hline
g & $-0.56 \pm 0.03$ & $4.26 \pm 0.08$ & 0.61 & 0.30  \\ \hline
\end{tabular}
$$
\centerline{ Table II}


\begin{thebibliography}{99}

\bibitem{pdf}
See for example H. Plothow-Besch; Computer Physics Comm. 75 (1993)  
396.

\bibitem{aversa}
F.Aversa, P.Chiappetta, M.Greco and  J.Ph.Guillet; Nucl.Phys.B327  
(1989) 105.


\bibitem{noi}
P. Chiappetta, M. Greco, J. Ph. Guillet, S. Rolli and M. Werlen;
Nucl. Phys. B 412 (1994) 3.

\bibitem{noieta}
M. Greco and S. Rolli; Z.Phys. C 60 (1993) 169.

\bibitem{fermi}
M. Greco , S. Rolli and A. Vicini;
Fermilab-Pub-94/075-A and DFPD 94/TH/21

\bibitem{copia}
J. Binnewies, B. A. Kniehl and G. Kramer;
Desy 94-124

\bibitem{lep}
Aleph Collaboration;
Contribution to the 27th International Conference on High Energy  
Physics, Glasgow, Scotland. 20-27 July 1994


\bibitem{franc}
F.M.Borzumati, B.A. Kniehl and G Kramer;
Z.Phys. C 57 (1993) 595

\bibitem{dg}
R. Baier, J. Engles and B. Petersson; Z. Phys. C 2 (1979) 265;
M. Anselmino, P. Kroll and E. Leader; Z. Phys. C 18 (1983) 307.

\bibitem{ellis}
R.K.Ellis and J.C.Sexton; Nucl.Phys.B 269 (1986) 445.

\bibitem{mt}
J.G.Morfin and Wu-Ki Tung; Z.Phys. C 52 (1991) 13.
     

\bibitem{mrs}
A. D. Martin, R. G. Roberts and W. J. Stirling; 

Phys. Rev. D 47 (1993) 867.
%Durham Preprint, DTP/92-16 (1992)

\bibitem{grvpro}
M. Gl\"uck, E. Reya and A. Vogt; Z.Phys. C 53 (1992) 127

\bibitem{herwig}
G. Marchesini, B.R. Webber; Nucl Phys. B 238 (1984) 1.

\bibitem{herwig2}
G. Marchesini, B.R. Webber; Nucl Phys. B 310(1988) 461.

\bibitem{isr}
F. B\"usser et al.; Nucl. Phys. B 106 (1976) 1.

\bibitem{e706}
E706 Collaboration; talk given at Moriond 94.


\bibitem {jets}
S.D. Ellis, Z. Kunszt and D.E Soper; Phys.Rev D40 (1989) 2188.

\bibitem{huth}
J.E.Huth et al., Fermilab-Conf-90/249-E (1990).

\bibitem{work}
Work in progress.


\bibitem{diph}
CDF Collaboration, Phys. Rev D (1993) 2998.

\bibitem{hubbard}
CDF Collaboration, Phys. Rev. Lett. (1990) 968.


\end{thebibliography}
\end{document}